\documentclass[aps,prl,twocolumn,showpacs,floatfix]{revtex4}
\usepackage{epsfig}
\usepackage{amssymb}

\newcommand{\ie}{{\it i.e.\ }}
\newcommand{\ignore}[1]{}
\begin{document}
\title{Statistics of cycles in large networks}
\author{Konstantin Klemm}
\author{Peter F. Stadler}
\affiliation{Dept. of Bioinformatics, University of Leipzig, H\"artelstrasse 16-18, D-04107 Leipzig, Germany} 
\date{\today} 
\begin{abstract} 
We present a Markov Chain Monte Carlo method for sampling cycle length in 
large graphs. Cycles are treated as microstates of a system with many degrees of
freedom. Cycle length corresponds to energy such that the length histogram
is obtained as the density of states from Metropolis sampling.
In many growing networks, mean cycle length increases algebraically with system
size. The cycle exponent $\alpha$ is characteristic of the local growth rules and
not determined by the degree exponent $\gamma$. For example, $\alpha=0.76(4)$ for
the Internet at the Autonomous Systems level.
\end{abstract} 
\maketitle  


Physics research into graphs and networks has begun to provide a common framework
for the analysis of complex systems in diverse areas including the Internet,
biochemistry of living cells, ecosystems, social communities
\cite{AlbertReview,DorogovtsevReview,BornholdtBook}. The graph representation of
these systems as discrete units coupled by links (nodes and edges) exhibits a
large set of scaling phenomena including fractal dimension \cite{Song05} and
hierarchy of modules \cite{Ravasz03}.

A fundamental observation is the {\em scale-free} nature of many networks
\cite{Barabasi99}. The fraction of nodes with a given
number of connections, called degree $k$, decays as a power law, $P(k) \sim
k^{-\gamma}$ for large $k$. For typical exponents $\gamma<3$, the highly
inhomogeneous density of connections can give rise to efficient information
transfer \cite{Pastor01a} and enhanced failure tolerance \cite{Albert00}. 

Beside the
degree distribution and node-node distances, the presence of {\em cycles} is
a relevant property of networks. A cycle is a closed, not self-intersecting
path. Initially, mainly cycles of the minimal length $h=3$ were considered since
high abundance of triangles is taken as a sign of a clustered structure
\cite{Watts98}. Longer cycles gained attention recently. 
Approximations for the system size scaling of the number $c(h)$ of cycles of
length $h$ have been derived for various types of artificial networks
\cite{Bianconi03,Vazquez05,Marinari04,BenNaim05,Bianconi05a}.
It has been speculated \cite{Rozenfeld05}
that for generic networks the distribution $c(h)$ becomes sharply
peaked in the limit of large networks, $N\rightarrow\infty$. For the position of
the peak, an algebraic growth has been conjectured
$\langle h \rangle \sim N^\alpha$ with an exponent $\alpha \le 1 $ as the
leading characteristic \cite{Rozenfeld05}.

Verification of these fundamental conjectures, validity checks of the analytical
approximations, and comparisons with real-world networks have been difficult so
far, since an efficient method for finding the cycle length
distribution of a given network has been lacking. Direct enumeration of all
cycles is feasible only for small networks because the number of cycles
increases exponentially with the number of nodes in most cases.
Approximation by efficient sampling appears
the only possibility to numerically investigate the cycle structure in the
general case. Taking a step in this direction, Rozenfeld and co-authors have
introduced a stochastic search for cycles \cite{Rozenfeld05} as self-avoiding
random walks on the network. Although the method allows for a quick scan of
cycles on small networks, larger systems cannot be treated as the probability of
finding a given cycle is strongly suppressed with growing cycle length.
Therefore we suggest an alternative method that does not involve random walks on
the network.


\begin{figure}[hbt]
\centerline{\epsfig{file=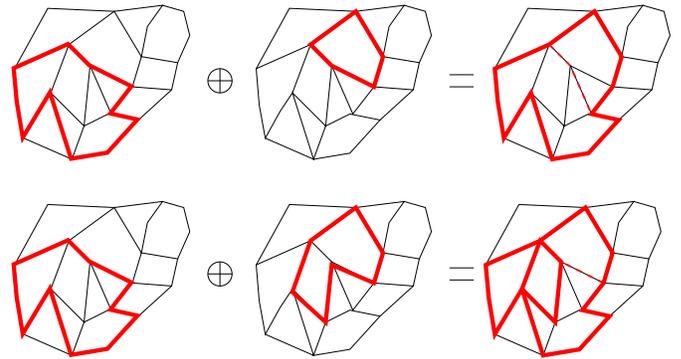,width=.49\textwidth}}
\caption{\label{fig:symdiff_illu}
(a) Summation of two cycles resulting in a new cycle. Edges contained in either addend are
contained in the sum. Edges present in both addends (dashed lines) cancel out.  
(b) Example of a sum of two cycles that is not a cycle itself. 
}
\end{figure}

We approximate the cycle length distribution by a Monte Carlo algorithm that
considers cycles as discrete microstates of a physical system. Elementary
transitions between cycles, the analogues of single spin flips in a spin system,
are defined as addition or removal of short detours with minimal change to cycle
length. By considering cycle length as energy, generic Monte Carlo procedures
from statistical mechanics become applicable. Temperature is defined in the
usual way and allows to tune the sampling on preferably long or short cycles.
%
%
After introducing the algorithm in detail, we test its accuracy for a set of
networks where the cycle length distribution is directly accessible for
comparison. We apply the algorithm to models of growing networks and find the
growth exponent of the mean cycle length. Finally, we test scaling of the number
of cycles in the growing Internet.


The formulation of the algorithm uses the following basic notions of cycle space.
We treat a subgraph $X$ as the set of edges it contains. If $X$ is a cycle, the
cardinality $|X|$ is the cycle length. The sum of two subgraphs $X$ and $Y$ is
defined as  $X \oplus Y = (X \cup Y) \setminus (X \cap Y)$, \ie an edge is contained
in the sum if it is in one of the addends but not in both. The sum $X \oplus Y$ of 
two cycles $X$ and $Y$ is again a cycle if $X$ and $Y$ intersect in a suitable way,
see Fig.\ \ref{fig:symdiff_illu}. We generate a
Markov chain of cycles $(C_0, C_1, C_2, \dots)$ as follows. The initial condition
is the empty graph $C_0=\emptyset$ at $t=0$. At each step a cycle $S$ is drawn at
random from a set $M$ of initially known cycles (the choice of $M$ is
described below). If the proposal $C^\prime = C_t \oplus S$ is a cycle or the empty
graph, it is accepted with probability
\begin{equation}
P_{\rm accept} = \min \{\exp[-\beta(|C^\prime| - |C_t|)], 1\} ~.
\end{equation}
In case of acceptance we set $C_{t+1} = C^\prime$, otherwise $C_{t+1}=C_t$.
This is the Metropolis update scheme \cite{Metropolis53} with inverse temperature
$\beta$ and energy as cycle length. Subgraphs that are not cycles are treated as
states with infinite energy $E=\infty$ if $\beta>0$ (or $E=-\infty$ if $\beta<0$,
respectively), such that they are always rejected.


Throughout this paper, we take $M$ as the set of {\em short} (isometric)
cycles of the given
graph. A cycle $S$ is short if for all vertices $x$ and $y$
on $S$, a shortest path between $x$ and $y$ lies also in $S$. As a non-short
cycle has at least one short-cut between two of its vertices, it can be 
decomposed into two shorter cycles that overlap on the short-cut. Typically for
each non-short cycle $C$ one finds cycles $S$ and $C^\prime$ such that $S$ is
short and $|C^\prime|<|C|$. Applying the decomposition recursively, one sees that
every cycle $C$ occurs in a sequence $0,C_1,C_2,\dots$ with $C_i \oplus C_{i+1}
\in M$ and $|C_i| < |C_{i+1}|$. Thus taking as the possible ``moves'' $M$ the set
of short cycles
not only ensures that every cycle can be reached (ergodicity). In this case,
the resulting
energy landscape does not have any local minima other than the unique global
minimum, which is the empty graph at $E=0$. There are exceptional
graphs where the decomposability does not hold for one particular cycle. The
exceptions appear to be irrelevant for the applications here as our numiercal results
remain unchanged when $M$ is expanded to include more and longer (non-short)
cycles.


\begin{figure}[hbt]
\centerline{\epsfig{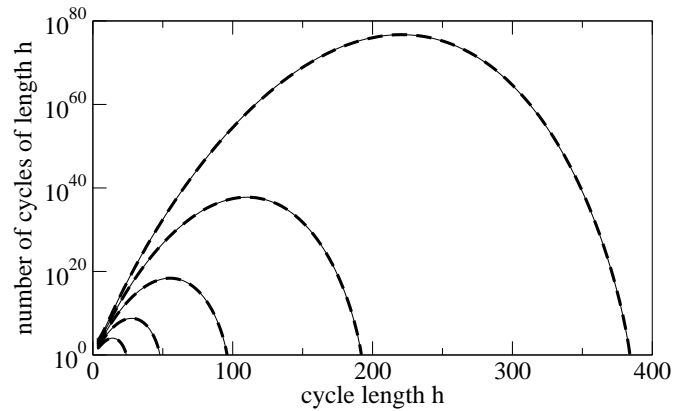}}
\caption{\label{fig:detsf} Number $c(h)$ of
cycles of length $h$ estimated by the MC sampling algorithm (thick dashed curves)
and the exact values from iterating Eq.\ \protect(\ref{eq:detsf_iterate}) (thin
solid curves). Studied networks are generations $n=4,\dots,8$ (system sizes $N=42,
123, 366, 1095, 3283$ vertices) from the deterministic growth model
\cite{DorogovtsevReview}. Given a network, a histogram is generated for each
inverse temperatures $\beta\in[-5.0,\dots,+3.0]$ in steps of $\Delta\beta=0.1$.
Each histogram is based on the lengths of the last $10^8$ cycles of a Markov chain
of total length $2\times 10^8$. Then histograms are merged by choosing relative
normalization such that the sum of squares of deviations in the overlapping region
of adjacent histograms are minimized. The normalization of the final histogram is
chosen such that $c(0)=1$. Results are robust against variation of the chain
length.
}
\end{figure}

Let us first test the algorithm
on a set of networks where exact computation of $c(h)$ is feasible.
The pseudo-fractal scale-free web by Dorogovtsev and Mendes
\cite{Dorogovtsev02} grows deterministically by iterative triangle
formation as follows. 
Start at generation $n=0$ with two vertices connected by an edge.
To obtain generation $n+1$, for each edge $xy$ present in
generation $n$ add a new vertex $z$ and the edges $xz$ and $yz$,
such that each existing edge $xy$ becomes part of an additional triangle
$xyz$. The calculation of $c(h)$ is particularly
simple because each cycle has a unique predecessor in the
previous generation, given by following direct links
$xy$ instead of the additional ``detours'' via $z$. A cycle of
length $h$ in generation $n$ produces $2^h$ cycles in generation
$n+1$ as the result of $h$ binary decisions to follow the detour
or the original direct edge. The histogram of cycle lengths
iterates as
\begin{equation} \label{eq:detsf_iterate}
c^{(n+1)}(h) = \sum_{l=3}^h {h \choose h-l} c^{(n)}(l)
\end{equation}
for $l\ge4$ and $c^{(n+1)}(3)=c^{(n)}(3)+3^n$. The result of the
numerical iteration of these equations up to generation $n=8$
is shown in Fig.\ \ref{fig:detsf}, together with the results
from the Monte Carlo method. The relative deviation of the sampling
estimate
of $c^{(n)}(h)$ from  the exact value is below $25\%$ for all
cycle lengths $h$ and all generations $n$. In particular, the
unique cycle of maximum length $h_{\rm max} = 3 \times 2^n$ is
detected. The method approximates the true numbers of cycles
with large precision.


\begin{figure}[hbt]
\centerline{\epsfig{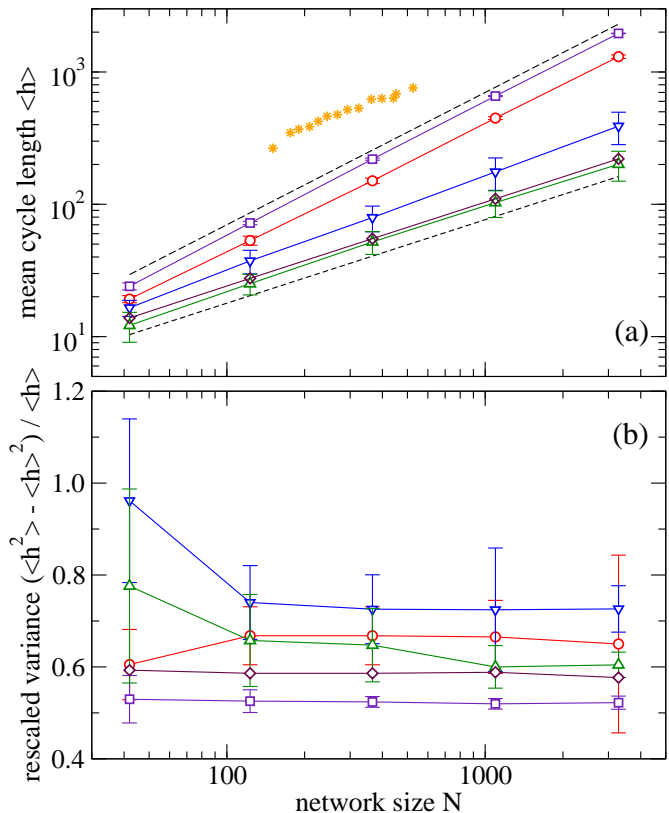}}
\caption{\label{fig:moments} System size dependence of the cycle length
distribution in growing networks. (a) Mean cycle length for
the four stochastic attachment rules ($\Box, \bigcirc, \nabla, \triangle$)
and the deterministic attachment rule ($\diamondsuit$).
For the Internet ($\ast$), system size $N$ has been rescaled by factor 20 to
fall into the displayed range. Dashed straight lines indicate growth exponents
1 and $\ln 2 / \ln 3 \approx 0.63$ for comparison.
(b) Relative variance of the cycle length distribution for the same networks
(same symbols as in (a)). In both panels, data points the
stochastic growth models are averages over 10 network realizations each.
Error bars indicate standard deviation over realizations.
}
\end{figure}

\begin{table}[hbt]
\caption{\label{table1} Networks with different attachment rules and the resulting
scaling exponents $\gamma$ for the tail of the degree distribution and $\alpha$ 
for the growth of the cycle lengths. The last column displays the symbol used in
Fig.\ \ref{fig:moments}.}
\begin{tabular}{|l|c|c|l|l|c|}
\hline
rule                    &indep / tri& hom / pref   & $\alpha  $ & $\gamma$ &           \\
\hline \hline
IH \cite{Barabasi99}   &independent& homogeneous  & $1.010(4)$ & $\infty$ & $\Box$\\ \hline
IP \cite{Barabasi99}   &independent& preferential & $0.969(5)$ & $3$      & $\bigcirc$\\ \hline
TH                      &triangle   & homogeneous  & $0.722(5)$ & $\infty$ & $\nabla$\\ \hline
TP \cite{Dorogovtsev01}&triangle   & preferential & $0.644(9)$ & $3$      & $\triangle$ \\ \hline
PF \cite{Dorogovtsev02}&triangle   & preferential & $0.635(1)$ & $2.59$   & $\diamondsuit$\\ \hline
\multicolumn{3}{|c|}{Internet}                      & $0.76(4) $ & $2.22(1)$& $\ast$\\ \hline   
\end{tabular}
\end{table}

Now we apply the algorithm to study the system size dependence of the cycle
length distribution of stochastically growing artificial networks. All networks
initiate as two vertices coupled by an edge. The networks grow by iterative
attachment of  vertices until a desired size $N$ is reached. At each iteration, one
new vertex $z$ and two new edges $xz$ and $yz$ are generated. We are interested
in the influence different attachment mechanisms have on the cycle length
distribution. Therefore we distinguish four probabilistic rules for selection of
the nodes $x$ and $y$ to which the new node $z$ attaches.
%
Independent homogeneous (IH) attachment: Draw $x$ and $y$ randomly (with equal
probabilities) and independently from the set of nodes; if $x=y$, discard this
choice and repeat.
Independent preferential (IP) attachment: Draw an edge randomly (all edges
having equal probability) and take as $x$ one of the end vertices chosen with
equal probability; draw another edge to find $y$ analogously; if $x=y$, discard
this choice and repeat.
Triangle forming preferential (TP) attachment: Draw an edge randomly and take
its two end vertices as $x$ and $y$. 
Triangle forming homogeneous (TH) attachment:
Draw an edge randomly, take $x$ and $y$ as its end vertices and accept this
choice with probability $1/(\deg (x) \deg(y))$; otherwise reject and repeat.

Rule IP is equivalent to choosing nodes with probability proportional to
degree \cite{Barabasi99}, so-called preferential attachment. It generates
scale-free networks with degree exponent $\gamma=3$.
Rule TP implements preferential attachment with the additional constraint
that $x$ and $y$ be connected; it is the stochastic version of the pseudo-fractal
(PF) scale-free web \cite{Dorogovtsev01} defined above.
The resulting networks are scale-free with $\gamma=3$. The homogeneous attachment
rule (IH) \cite{Barabasi99} leads to networks with exponentially
decaying degree distribution ($\gamma=\infty$). The fourth rule (TH) introduced
here combines triangle formation with homogeneous attachment by explicitly
canceling out the degree dependence in the selection probability. We have
checked that this rule generates an exponential degree distribution.

As shown in Fig.\ \ref{fig:moments}(a) the mean cycle length increases
algebraically with system size,
\begin{equation}
\langle h \rangle \sim N^\alpha~,
\end{equation}
with the exponent $\alpha\in[0,1]$ depending on the attachment rule.
The variance of the cycle length distribution increases algebraically
with the same exponent $\alpha$. Therefore the ratio between variance
and mean is practically constant, see Fig.\ \ref{fig:moments}(b).
\ignore{
\begin{equation}
\langle h^2 \rangle - {\langle h \rangle}^2 \sim \langle h \rangle~.
\end{equation}
}
Considering the degree exponent $\gamma$ and the cycle growth exponent
$\alpha$ for each type of network (Table \ref{table1}), several observations
are worth mentioning. Homogeneous attachment with triangle formation
leads to a non-trivial cycle growth exponent $\alpha\approx 0.72$ even
in the absence of scaling in the degree distribution $\gamma=\infty$.
Networks grown stochastically with triangle formation and preferential
attachment (rule TP) have the same exponent $\alpha\approx0.64$
as the deterministic counterpart (rule PF) while the degree exponents
under these two rules are clearly different. Analogously, in the absence
of triangle formation (rules IH and IP) the same cycle growth exponent
$\alpha\approx 1.0$ is obtained regardless of the degree exponents
$\gamma \in \{3,\infty\}$. 


\begin{figure}[hbt]
\centerline{\epsfig{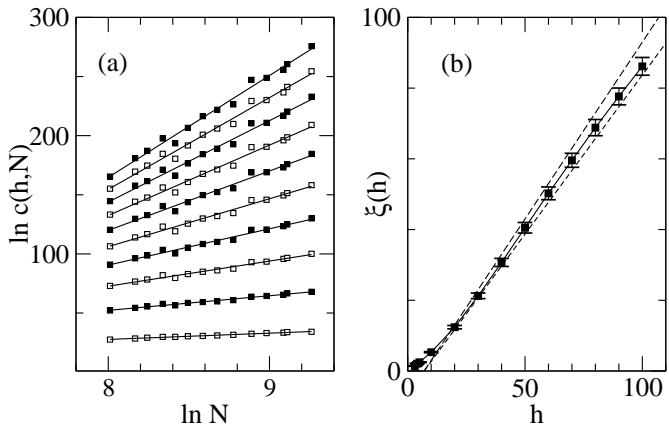}}
\caption{\label{fig:AS_0} Evolution of cycles in the growing Internet at the 
Autonomous Systems level. (a) The number of cycles of given length $h$ as
a function of system size $N$ for $h = 10,20,30,\dots,100$ (squares,
bottom to top). The straight lines are best fits of the form
$c(h,N) \propto N^{\xi(h)}$.
(b) Growth exponents $\xi(h)$ as defined in Eq.\ (\ref{eq:xi}) obtained
as slopes of the fitted lines in (a).  Error bars of exponents indicate
standard error from the fit. Dashed lines have slopes $1.0$ and $0.9$.
}
\end{figure}

Finally we consider cycles in an evolving real-world network.
The Internet at the level of Autonomous Systems is a growing scale-free
network with degree exponent $\gamma=2.22(1)$ \cite{Faloutsos99,Pastor01b}.
Here we analyze snapshots of the network with sizes from $N=3015$
nodes (November 1998) to $N=10515$ nodes (March 2001) \cite{InternetData}.
We find that during this time the mean cycle length grows from 264.9 to
757.8, as plotted in Fig.\ \ref{fig:moments}(a). As in the artificial growing
networks, the growth is algebraic. The growth exponent is estimated as
$\alpha = 0.76(4)$ by a least squares fit. More detailed analysis is performed
on the number $c(h,N)$ of cycles of given length $h$ at
system size $N$ plotted in Fig.\ \ref{fig:AS_0}(a). We observe a scaling
\begin{equation} \label{eq:xi}
c(h,N) \sim N^{\xi(h)}~.
\end{equation}
with an exponent $\xi(h)$ that depends linearly on $h$ with a slope close to
unity. Figure \ref{fig:AS_0}(b) shows that
\begin{equation}
\xi(h) \approx h~.
\end{equation} 
for not too small lengths $h \ge 10$.
The scaling behavior is in qualitative agreement with the prediction from
the first order approximation by Bianconi et al.\ \cite{Bianconi05b}, assuming
that the Internet is a random network with a given scale-free degree distribution.

In summary, we have introduced a method for sampling cycles in large graphs.
We have identified cycle space with the state space of a system with
many degrees of freedom, thereby making Monte Carlo techniques from statistical
mechanics applicable. In this framework, we have analyzed the evolution of
cycles in growing networks. While the mean cycle length grows with a
characteristic exponent $\alpha$ the relative width of the length distribution
tends to zero as the system size increases. Thus, in agreement with an earlier
speculation \cite{Rozenfeld05}, the exponent $\alpha$ is found to be the most
relevant quantity for the evolution of cycle space. In the scale-free model by
Barab\'asi and Albert \cite{Barabasi99} as well as the growth model with random
homogeneous attachment, cycles are space-filling ($\alpha=1.0$), \ie cycle
length is proportional to system size. In model networks with explicit formation
of triangles and in the Internet, however, cycles grow slower than the system
as a whole. This class of networks having $\alpha<1$ also includes single-scale
networks
with $\gamma=\infty$. Our study suggests that the cycle growth exponent may
serve as a characterization of growing networks independent of the degree
exponent $\gamma$. An open question concerns universality. Can $\alpha$ be
altered continuously by tuning parameters or does it assume distinct values,
separating growing networks into universality classes?

We are grateful to C. P. Bonnington, J. Leydold, and A. Mosig for inspiring
discussions. This work was supported by the DFG Bioinformatics Initiative
BIZ-6/1-2.


\end{document}